\documentclass{aa}  
\usepackage{graphicx}
\usepackage{txfonts}
\usepackage[colorlinks=true, citecolor=blue, linkcolor=blue]{hyperref}
\usepackage{xcolor}
\usepackage[normalem]{ulem}
\usepackage{natbib}
\bibpunct{(}{)}{;}{a}{}{,} 

\begin{document}

   \title{Properties of granular light bridges}

   \subtitle{}

   \author{Michal Sobotka 
          \and
          Jan Jur\v{c}\'{a}k 
          \and
          Marta Garc\'\i a-Rivas 
          }

   \institute{Astronomical Institute of the Czech Academy of Sciences,
              Fri\v{c}ova 298, 25165 Ond\v{r}ejov, Czech Republic \\
              e-mail: {\tt michal.sobotka@asu.cas.cz}
    }

  \date{Received 20 May 2026; accepted 4 August 2026}

\abstract
   {Granular light bridges in sunspots are a manifestation of magnetoconvection in an approximately vertical magnetic field that is weaker than the field of the umbra.}
   {We aim to study the relations between observed properties of granules and magnetic field in five strong granular light bridges located in a complex sunspot.}
   {Thermal parameters, magnetic field vector, and line-of-sight velocity were obtained from an inversion of the GREGOR spectropolarimetric scan in the lines \ion{Si}{i} 1082.71\,nm and \ion{Ca}{i} 1083.90\,nm. Sizes and lifetimes of granules were measured in a simultaneous sequence of images in the TiO band.}
  {The light bridges under study are comprised of granules spanning 0\farcs 2–-1\farcs 4 with approximately 6-minute lifetimes. The largest granular structures ($> 1$\arcsec) are spatially correlated with 1~km\,s$^{-1}$ upflows and weak, inclined magnetic fields ($\simeq 300$~G) near the equipartition threshold. Statistical comparisons demonstrate an inverse relationship between magnetic field strength and continuum intensities, line-of-sight velocity dispersions, and effective granular diameters. These relationships vary among individual light bridges, depending on the large-scale magnetic field topology.}
   {Large expanding granules similar to the photospheric ones appear in light bridges only when the local magnetic field strength approaches the equipartition value.}


  \titlerunning{Granular light bridges}

\maketitle
\nolinenumbers   
%

\section{Introduction}
\label{sec:intro}

Light bridges (LBs), bright elongated structures that cross the sunspot umbra or deeply penetrate into it, are important elements of the structure of sunspots. Their lifetime can reach several days but their shape may change substantially on the scale of hours. Light bridges show a large variety of sizes, brightnesses, and shapes. Several authors attempted to classify LBs according to their morphology \citep{Bray64, Muller79, Sob97, ThW04, Lagg14, Esteban24}. The Sobotka's classification is based on two parameters: The first one is the morphology related to the sunspot configuration -- if the LB separates umbral cores, it is called strong and if it is an intrusion into the umbra, it is called faint. The second parameter is the LB's internal structure, which can be granular or filamentary.

Strong granular LBs, located between two or more umbral cores, are relatively bright, comparable with the penumbral or even the quiet Sun brightness, and they are filled by granules of different sizes \citep[e.g.,][]{SBV94, Berger03}. Faint granular LBs, sometimes called umbral LBs, are narrower and less bright than the strong ones. They are embedded in a single umbral core and consist of small bright grains similar to umbral dots \citep[e.g.,][]{SBV93, Katsukawa07}. Strong filamentary LBs \citep{Bumba80, Ruedi95, Qiao26} resemble distorted parts of a penumbra that separate umbral cores. Faint filamentary LBs develop as intrusions of penumbral filaments, including moving elongated bright grains. A combination of filamentary and granular structures in a single LB or a transition from one type of the structure to another \citep{Shimizu11, Song24} has also been observed. Recently, \citet{Schli16} reported a new type, the plateau LB, which is as broad as strong granular LBs but it is flat, lacking the granular structure.

The evolution of LBs is related to the development of sunspots. During sunspot formation, areas of photospheric granulation or penumbra compressed between merging umbrae develop into LBs; the width and brightness of these decrease and during further evolution they may disappear. A reverse process can take place during the sunspot life, when faint LBs increase the width and brightness until they split the umbra in the form of strong LBs \citep[e.g.,][]{Grinon21}. Finally, when the spot decays, strong granular LBs develop into regions of photospheric granulation between fragments of the umbra \citep{Vazquez73}.

\begin{figure*}[t]\centering
\includegraphics[width=0.99\textwidth]{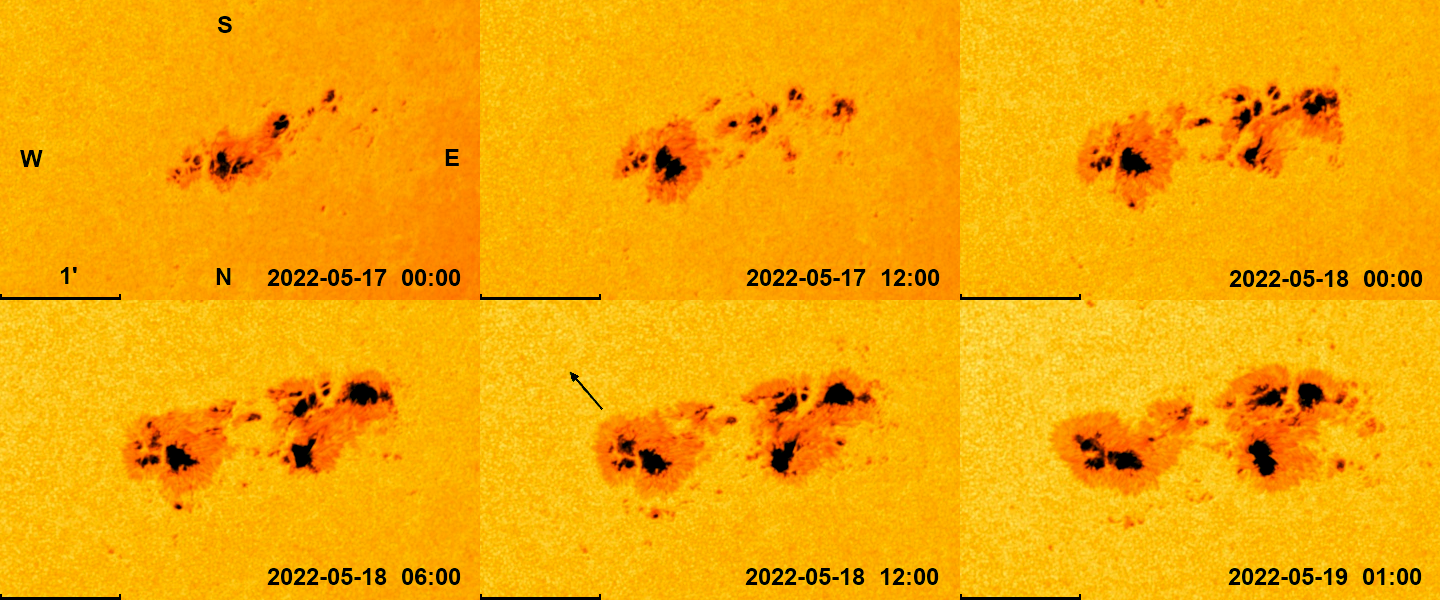}
\caption{Temporal evolution of AR 13014 observed in the HMI continuum. The orientation is equal to that of GREGOR observations and the leading spot is on the left side. The line segments are 1 minute-of-arc long. The arrow shows the direction to the solar disc centre at the time of observations.
\label{fig:hmi}}
\end{figure*}

Many granular LBs show a narrow dark lane running along their main axis when observed with high angular resolution \citep[e.g.,][]{Berger03, Giordano08, Rouppe10}. The dark lanes can be explained by the accumulation of rising plasma at the top of the LB structure, similarly to the case of dark cores of bright penumbral filaments \citep{Ruiz08}.

All authors agree that the magnetic field is weaker and more horizontal in photospheric layers of LBs than in the surrounding umbra \citep[e.g.,][]{Leka97, Jurcak06, Felipe16, Falco16, Grinon21, Esteban24}. \citet{Leka97} first proposed presence of a magnetic canopy structure above LBs and \citet{Jurcak06} confirmed this model, deriving the vertical stratification of plasma parameters with optical depth. The magnetic field strength is weak in deep layers of LBs and the area filled with the weak-field plasma is contracting with height. The magnetic inclination decreases with height and a magnetic canopy extending from either side of the LB is formed. The field lines from both sides then meet and become more vertical above the LB, where the magnetic field strength is comparable with that of the surrounding umbra. However, the magnetic canopy is not observed in all LBs \citep{Esteban24} and never in bi-polar LBs that separate umbrae of opposite polarity \citep{Castellanos:2025}.

The velocity field measured in light bridges shows blueshifts and redshifts of the order of 1~km\,s$^{-1}$ \citep{Leka97, Rimmele97, Giordano08, Rouppe10, Toriumi15}. \citet{Rimmele97} found a clear anticorrelation between the continuum intensity and line-of-sight (LOS) velocity (when upflows have negative values) in a strong granular LB, which indicates the magnetoconvective origin of this type of LB. \citet{Louis09} found supersonic downflows reaching 10~km\,s$^{-1}$ in small patches of a strong LB. Later, \citet{Lagg14} reported central hot upflows surrounded by cool supersonic downflows in granules forming a strong granular LB. From the similarity of LB, plage, and quiet-Sun granules they confirmed that these flows are produced by magnetoconvection and that strong granular LBs are anchored in deep layers.

Granular LBs are a manifestation of magnetoconvection in approximately vertical magnetic field. Strong granular LBs are related to intrusions of weak-field or possibly field-free plasma from deep convective layers below the sunspot. On the visible surface, the magnetoconvection produces granules of various sizes. According to \citet{SBV94}, the typical size of large LB granules is
1\farcs 2, slightly smaller than the quiet-Sun granules (1\farcs 5). In addition to these, numerous small grains of 0\farcs 5 in size are observed. This work aims to study the relations between observed properties of granules and magnetic field vector in five strong granular LBs located in a complex sunspot.

\section{Observations and data processing}
\label{sec:data}

The active region (AR) 13014 appeared on 2022 May 16. Its growth phase was monitored in continuum images (Fig.~\ref{fig:hmi}) provided by the Helioseismic and Magnetic Imager \citep[HMI,][]{Scherrer12, Schou12} on board the Solar Dynamics Observatory \citep[SDO,][]{Pesnell12}. AR 13014 grew rapidly, forming first the leading part and later, on May 17 between 12:00 and 18:00 UT, the following part. The sunspot group reached its maximum area on May 18 and 19. On May 17, the leading spot developed a large umbra accompanied on the west by three small spots of equal magnetic polarity with partial penumbrae. These approached the main umbra on May 17 around 12:00 UT and started to form a large spot studied in this work. The granulation between the western and main parts was squeezed in LBs. The joint penumbra of the spot developed between 00:00 and 06:00 UT on May 18 and at the same time appeared a new umbral core south-west from the main umbra. Two and half hours later, during our observations, five strong granular LBs were present inside the sunspot. 

The leading sunspot of AR 13014 was observed with the GREGOR 1.5-m solar telescope \citep{Schmidt12} on 2022 May 18 at the position $-375$\arcsec E, 390\arcsec N, heliocentric angle $\vartheta = 35$\degr, $\mu = \cos \vartheta = 0.82$. It was scanned with the GREGOR Infrared Spectropolarimeter \citep[GRIS,][]{Collados12} from 08:29 to 08:48 UT with a spatial sampling of 0\farcs 135 per pixel. The spectral region contained lines \ion{Si}{i} 1082.71\,nm and \ion{Ca}{i} 1083.90\,nm used for further analysis. The spectral sampling was 1.81\,pm.
The correction of the GRIS data for dark- and flat-fields and the polarimetric calibration \citep{Collados99, Collados03} was performed using the standard GRIS data reduction software.

Profiles of the two spectral lines (\ion{Si}{i} 1082.71\,nm and \ion{Ca}{i} 1083.90\,nm) from the scan with the field of view (FOV) of 65\arcsec $\times$ 40\farcs 5 ($481 \times 300$ pixels) covering the whole leading sunspot were inverted simultaneously using the code Stokes inversion based on the response functions \citep[SIR,][]{Ruiz92}.
Although the core of the strong \ion{Si}{i} 1082.71\,nm line is formed outside the local temperature equilibrium, it does not affect the inversion results in deep photospheric layers near the visible surface.
We used two nodes for the microturbulence, LOS velocity, magnetic field strength, and inclination. The resulting linear stratification with optical depth makes it possible to use SIR results at the visible surface ($\log \tau_{500\rm nm} = 0$), because the linear parameters of the stratification are determined mainly at optical depths where the line profiles are most sensitive to physical conditions and not at the extreme positions of the nodes \citep[cf. the response functions by][]{Felipe16}. Three nodes were used to determine the temperature. Optical depths of the nodes are listed in Table~\ref{tab:nod}. The magnetic field azimuth was assumed to be constant with the optical depth. The 180\degr ~azimuthal ambiguity was resolved assuming radial orientation of the magnetic field in the penumbral filaments and the values of the magnetic field inclination and azimuth were transformed from the LOS frame to the local reference frame (LRF) using routines from the AZAM code \citep{Lites95}. The LOS velocities were calibrated by setting to zero the area-averaged velocity inside the largest umbra.

\begin{table}\centering
\caption{Optical depths, $\log \tau_{500\rm nm}$, used in the inversion}  \label{tab:nod}
\begin{tabular}{rrrrr}
\hline\hline
   $T$     & $v_{\rm mic}$ & $v_{\rm LOS}$ &  $B$   & $\gamma$ \\
\hline
   $-3.8$  &    $-3.8$     &    $-3.8$     & $-3.8$ &  $-3.8$  \\
   $-1.9$  &      -        &      -        &   -    &    -     \\
     1.0   &     1.0       &     1.0       &  1.0   &   1.0    \\
\hline
\end{tabular}
\tablefoot{$T$ -- temperature, $v_{\rm mic}$ -- microturbulent velocity, $v_{\rm LOS}$ -- line-of-sight velocity, $B$ -- magnetic field strength, $\gamma$ -- inclination.}
\end{table}

A series of broadband images in the TiO band at $750.7 \pm 1.0$\,nm was acquired simultaneously with the GRIS scan using the Improved High-resolution Fast Imager \citep[HiFI+,][]{Denker23}. Its spatial sampling was 0\farcs 05 per pixel.
The TiO images were processed by the sTOOLS software package \citep{Kuckein17stools}, using a frame selection routine \citep{Denker18} to select the best-quality image of each 500 frames taken in a time period of 5.5\,s. In total, 200 best images were selected from the observation sequence with the cadence of 5.47\,s. The selected images were aligned and destretched. Oscillations and a residual jitter caused by the seeing were removed by a subsonic filter with a cutoff at 4~km\,s$^{-1}$. The resulting filtered sequence with FOV 76\farcs 8 $\times$ 60\farcs 4 contained 184 frames acquired from 08:30:05 to 08:46:46 UT (duration 16\,m 41\,s) with a typical angular resolution of 0\farcs 3--0\farcs 4. A movie with a contrast enhanced by unsharp masking and marked frame numbers is available online.
The GRIS continuum map was resampled by linear interpolation to the scale of 0\farcs 05 per pixel and aligned with the average image of the TiO sequence. The alignment parameters (position, magnification, and rotation) were then used to align the maps of physical parameters, obtained from the inversion, with the TiO images. Finally, a working field of \mbox{19\farcs 5 $\times$ 20\farcs 5} covering all the LBs under study was adopted.

The broadband TiO images were segmented to isolate LB granules and measure their sizes and lifetimes. An algorithm based on a search for regions with convex intensity profiles was applied to the series of images with a contrast enhanced by unsharp masking. The resulting binary segmentation mask \mbox{(0: background, 1: granule)} was then refined using morphologic operators {\it close} to fill holes and gaps and {\it open} to remove noise and smooth edges. Sizes of granules are represented by the effective diameter $d_{\rm eff} = 0$\farcs 05$\sqrt{4A/\pi}$, where $A$ is a number of pixels that form a segmented granule.
Because the image stability did not allow to reliably track granules in time, lifetimes of granules were estimated by adding values of segmentation masks (0 or 1) at a given location, counting how many times a given pixel belongs to a segmented granule. This method ignores an intermittent appearance of a granule at the same position and gives a total period of presence ($t_{\rm p}$) of a granule, which can be used as a proxy of the lifetime.
Horizontal motions of intensity structures in LBs were estimated using local correlation tracking \citep[LCT,][]{November88}. We applied the LCT algorithm with a tracking window of 0\farcs 5 to the series of TiO images and averaged the flow maps over the whole series.

\section{Results}
\label{sec:results}

\begin{figure}[]\centering
\includegraphics[width=0.49\textwidth]{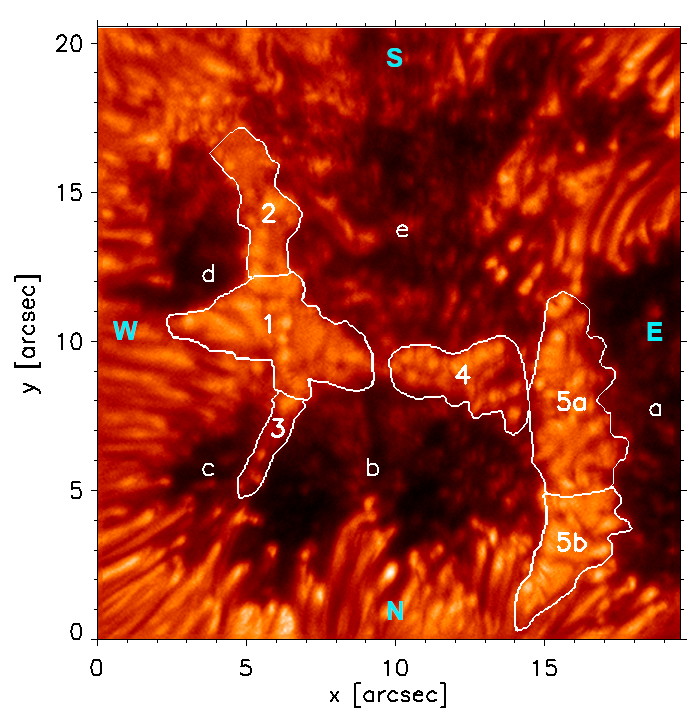}
\caption{Identification of light bridges (1--5a,b) and umbral cores ({\it a--e}) in a broadband TiO image taken on May 18, 2022, 08:43 UT. Symbols N, S, E, W show the orientation. The associated movie is available online.
\label{fig:bridges}}
\end{figure}

Five strong granular LBs 1--5 were identified in the analysed field of view and are displayed, together with five surrounding umbral cores {\it a--e}, in Fig.~\ref{fig:bridges}.
The borders of LBs were defined  primarily visually in the time-averaged image of the TiO series. Possible ambiguities, for example, a separation of LB~1 from LB~4 and the penumbra, or a division of LB~5 into two parts, were resolved using maps of magnetic field strength and inclination from the inversion results.
A single TiO frame is shown in Fig.~\ref{fig:bridges} to demonstrate the intensity structures in detail, therefore, the LB borders do not exactly match the intensity gradients in this frame.
An overview of LB formation and evolution, based on the HMI continuum images between May 17 and 19 (Fig.~\ref{fig:hmi}), is listed below:

\begin{figure*}[]\centering
\includegraphics[width=\textwidth]{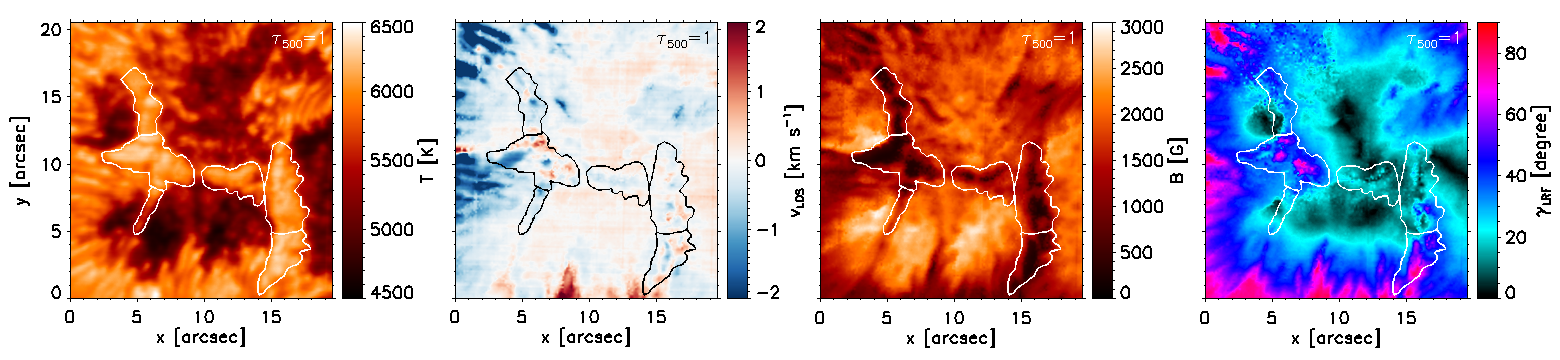}
\caption{Maps of temperature ($T$), line-of-sight velocity ($v_{\rm LOS}$), magnetic field strength ($B$), and magnetic field inclination in the local reference frame ($\gamma_{\rm LRF}$) at optical depth $\log \tau_{500\rm nm} = 0$. Contours outline the LB borders.
\label{fig:maps}}
\end{figure*}

LB 1 -- This broad LB was formed together with LBs~2 and 4 from a granular area between the umbral cores {\it b, c, d}, and later {\it e} on May 18 around 06:00 UT. It was visible until May 19 around 19:00 UT, when it separated umbral cores {\it e} and a merger of {\it a}+{\it b}. During the GREGOR observations (08:28--08:48 UT), LBs 2, 3, and 4 were adjacent to LB 1 and several large granules were formed in its western (Fig.~\ref{fig:bridges}) and also eastern parts.

LB 2 -- It was formed from a granular area between the umbral cores {\it d} and {\it e} on May 18. It lived approximately from 06:00 UT when {\it e} was formed to 18:00 UT when {\it d} disappeared. A central channel was present in this LB.

LB 3 -- This narrow chain of small granules was formed from a penumbral area between the umbral cores {\it b} and {\it c} on May 18 around 00:00 UT. It got broader and brighter at about 12:00 UT and disappeared at the beginning of May 19, when {\it b} and {\it c} merged.

LB 4 -- This LB, composed of small granules, was formed from granular and penumbral areas on May 18 around 06:00 UT together with the umbral core {\it e}. It was squeezed between {\it b} and~{\it e} and later merged with LB 1 at the beginning of May 19. A central channel was present in LB 4.

LB 5 -- This broad LB was formed on May 18 around 00:00 UT (before the umbral core {\it e} appeared) from a granular area between the largest umbral core {\it a} and the western core {\it b}. At the time of GREGOR observations, it separated umbral cores {\it a, b}, and {\it e}. Its southern part (LB 5a) showed a central channel and it was in contact with LB 4. Some large granules were formed in its northern part (LB 5b), which penetrated deeply into the neighbouring penumbra. Because the parts LB 5a and 5b had different magnetic field strengths and inclinations, we analyse them individually.

\subsection{Magnetic and velocity fields}
\label{sec:fields}

Results of the inversion, maps of temperature ($T$), LOS velocity ($v_{\rm LOS}$), magnetic field strength ($B$) and magnetic field inclination ($\gamma$) in LRF at the visible surface ($\log \tau_{500\rm nm} = 0$) are shown in Fig.~\ref{fig:maps}. We can see from the figure that the LB temperature is comparable to the penumbral one, with an exception of the narrow LB~3 where it is lower. The LOS velocity shows upflows (negative values) often accompanied by downflows (positive) in all LBs. The strongest ones, in the range $\pm 1.5$~km\,s$^{-1}$, are observed in LB~1. The upflows are mostly concentrated along the axes of LBs. The magnetic field strength is clearly reduced in LBs and the inclination depends on the geometry of surrounding strong fields but it is usually more horizontal in LBs than in the surroundings. 

\begin{table}\centering
\caption{Area-averaged parameters of individual light bridges}  \label{tab:ave}
\begin{tabular}{lrrrrrr}
\hline\hline
                             &  LB 1  &  LB 2  &  LB 3  &  LB 4  &  LB 5a  &  LB 5b \\
\hline
 $I_{\rm c}$ [$I_{\rm ph}$]  &  0.93  &  0.91  &  0.76  &  0.84  &  0.89   &  0.94 \\
 ~~$\sigma(I_{\rm c})$       &  0.05	&  0.04	  &  0.08  &  0.08  &  0.07	  &  0.05 \\
 $T$ [K]                     & 6030   & 6000   & 5580   & 5800   & 5950    & 6080  \\
 ~~$\sigma(T)$               &  160   &  110   &  230   &  230   &  190    &  140  \\
 $B$ [G]                     &  870   &  990   & 1620   & 1420   & 1150    & 860   \\
 ~~$\sigma(B)$               &  500   &  440   &  360   &  340   &  440    & 440   \\
$\gamma_{\rm LRF}$ [deg]     &   42   & 29     &   24   &   14   &   20    &  48   \\
~~$\sigma(\gamma_{\rm LRF})$ &   10   & 11     &    7   &    5   &    9    &  14   \\
$v_{\rm LOS}$ [m/s]          & $-200$ & $-310$ & $-340$ & $-70$  & $-50$   & $-20$ \\
~~$\sigma(v_{\rm LOS})$      &   560  &  290   &  270   &  200   &  200    &  320  \\
$d_{\rm eff}$ [\arcsec]      &  0.51  &  0.44  & 0.45   & 0.43	 &  0.43   &  0.50 \\
~~$\sigma(d_{\rm eff})$      &  0.26  & 0.17   &  0.17	& 0.16	 & 0.17    &  0.21 \\
$t_{\rm p}$ [min]         &  6.4	  & 6.0	   & 6.7	&  6.2	 &  5.5	   &  6.3  \\
~~$\sigma(t_{\rm p})$     &  3.9	  & 3.7	   &  4.7	& 4.6	 & 4.1	   &  4.3  \\
\hline
\end{tabular}
\tablefoot{$I_{\rm c}$ -- continuum intensity at $\lambda = 1084$\,nm in units of the local mean quiet photospheric intensity $I_{\rm ph}$. $T$ -- temperature, $B$ -- magnetic field strength, $\gamma_{\rm LRF}$ -- LRF inclination, $v_{\rm LOS}$ -- LOS velocity at $\log \tau_{500\rm nm} = 0$; $d_{\rm eff}$ and $t_{\rm p}$ are effective diameters and periods of presence (lifetimes) of granules, respectively. Scatter of individual values is given by a standard deviation $\sigma$ for each parameter.}
\end{table}

\begin{figure*}[]\centering
\includegraphics[width=\textwidth]{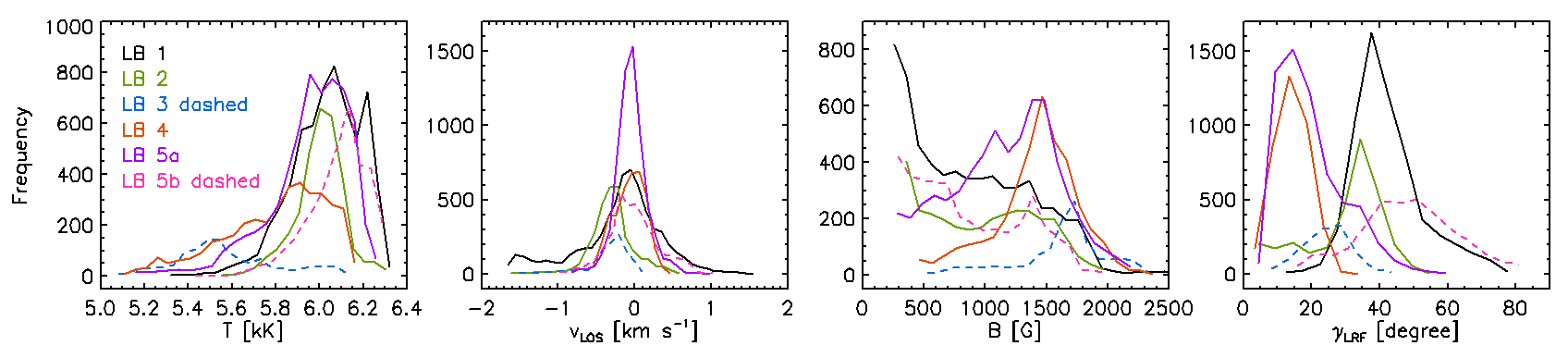}
\caption{Histograms of temperature ($T$), line-of-sight velocity ($v_{\rm LOS}$), magnetic field strength ($B$), and magnetic field inclination ($\gamma_{\rm LRF}$) in individual LB 1--5b areas at optical depth $\log \tau_{500\rm nm} = 0$.
\label{fig:hist1}}
\end{figure*}

Basic magnetic, velocity, and granular parameters averaged over individual areas of LBs 1--5b are summarised in Table~\ref{tab:ave}. The continuum intensity $I_{\rm c}$ is given in units of the mean local quiet-Sun continuum $I_{\rm ph}$ at $\lambda = 1084$\,nm. The temperature, LOS velocity, and magnetic field strength and inclination at \mbox{$\log \tau_{500\rm nm} = 0$} are obtained from the inversions. Granular sizes and periods of presence (a proxy of lifetimes) are measured in the TiO band (Sect.~\ref{sec:gran}). Scatter of values around the average is characterised by a standard deviation $\sigma$ for each parameter. We can distinguish between LBs with a weak magnetic field ($B < 1000$\,G: LBs 1, 2, 5b), a strong one ($B > 1000$\,G: LBs 3, 4, 5a), a nearly vertical field ($\gamma \leq 20$\degr: LBs 4, 5a), a moderately inclined one ($\gamma \simeq 27$\degr: LBs 2, 3), and a more inclined one ($\gamma > 40$\degr: LBs 1, 5b).

Histograms of $T$, $v_{\rm LOS}$, $B$, and $\gamma_{\rm LRF}$ at $\log \tau_{500\rm nm} = 0$ are shown for each LB area in Fig.~\ref{fig:hist1}. Typical temperatures in most LBs are in the range 5800--6300~K. The highest temperatures of 6300~K are observed in LBs~1, 5b, and a small part of LB~2 with weak magnetic fields. Bridges located in a strong magnetic field show low-temperature tails down to 5100~K (LBs~3, 4, and 5a) and LB~3 is the coolest one with a typical temperature around 5500~K. The LOS velocities show slightly prevailing downflows but their detailed distributions are different in individual LBs. They are approximately symmetric within the range of $\pm 0.7$~km\,s$^{-1}$ in LBs 4 and 5a, while LB 5b has a slightly asymmetric distribution in the range of $\pm 0.8$~km\,s$^{-1}$. Upflows dominate in LBs 2 and 3 in the ranges of $-1.0$--0.5~km\,s$^{-1}$ and $-1.0$--0.1~km\,s$^{-1}$, respectively. The strongest upflows ($-1.7$~km\,s$^{-1}$) and downflows (1.2~km\,s$^{-1}$) are observed in LB 1 but the typical value there is $-0.1$~km\,s$^{-1}$. The magnetic field strength ranges from 300~G (LBs 1, 2, 5a, 5b) or 500~G (LBs 3, 4) to approximately 2300~G (all LBs). Although the range of values is approximately equal in all LBs, the histograms differ substantially, showing absolute maxima at approximately 300~G (LBs 1, 2, 5b), 1500~G (LBs 4, 5a), and 1700~G (LB 3). Histograms of the LRF inclination illustrate the three above-mentioned LB groups with a nearly vertical field (LBs~4 and 5a), a moderate inclination (LBs~2 and 3), and an inclination above 40\degr (LBs~1 and 5b).

\begin{figure}[]\centering
\includegraphics[width=0.49\textwidth]{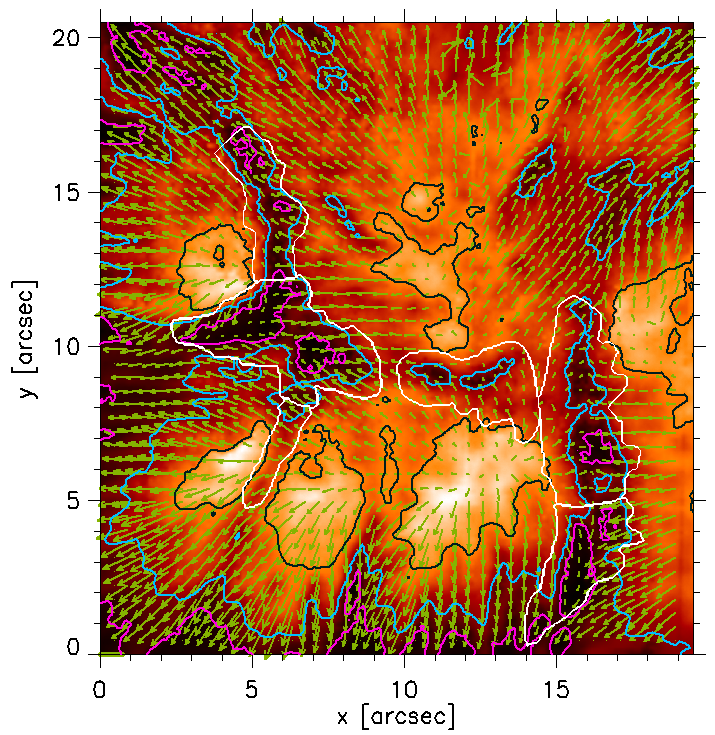}
\caption{Map of vertical ($B_{\rm z}$) and horizontal ($\vec B_{\rm h}$) components of the magnetic field vector. The colour coding of $B_{\rm z}$ ranges from 0 (black) to 2.7~kG (white) and the magenta, blue, and black contours display $B_{\rm z} = $~0.3, 1.0, and 2.0~kG, respectively. Green arrows represent $\vec B_{\rm h}$; a green segment in the lower left corner corresponds to $B_{\rm h} = 1$~kG. White contours delimit the LB areas.
\label{fig:bzbh}}
\end{figure}

A detailed magnetic field configuration at $\log \tau_{500\rm nm} = 0$ is shown in a map of the vertical ($B_{\rm z}$) and horizontal ($\vec B_{\rm h}$) magnetic components, depicted in Fig.~\ref{fig:bzbh}. The red colour scale of $B_{\rm z}$ ranges from 0~G (black) to 2700~G (white) and the magenta, blue, and black contours display the values of 300, 1000, and 2000~G, respectively. There are several large areas of $B_{\rm z} < 300$~G in LBs 1, 5a, and 5b. Central channels in LBs 2, 4, and 5a are co-spatial with $B_{\rm z} < 1000$~G and also with upflow areas (cf. Fig.~\ref{fig:maps}). The horizontal-component vector $\vec B_{\rm h}$ is represented by green arrows. Its azimuthal orientation shows that the magnetic inclination inside LBs is determined by the magnetic field of neighbouring umbral cores, which fans out with height. Magnetic fields of opposing umbral cores may form a magnetic canopy above the western part of LB 1 located between the cores {\it c} and {\it d}, above LB 4 between the cores {\it b} and {\it e}, and above LB 5a between the cores {\it a} and {\it b, e}.

\begin{figure}[]\centering
\includegraphics[width=0.49\textwidth]{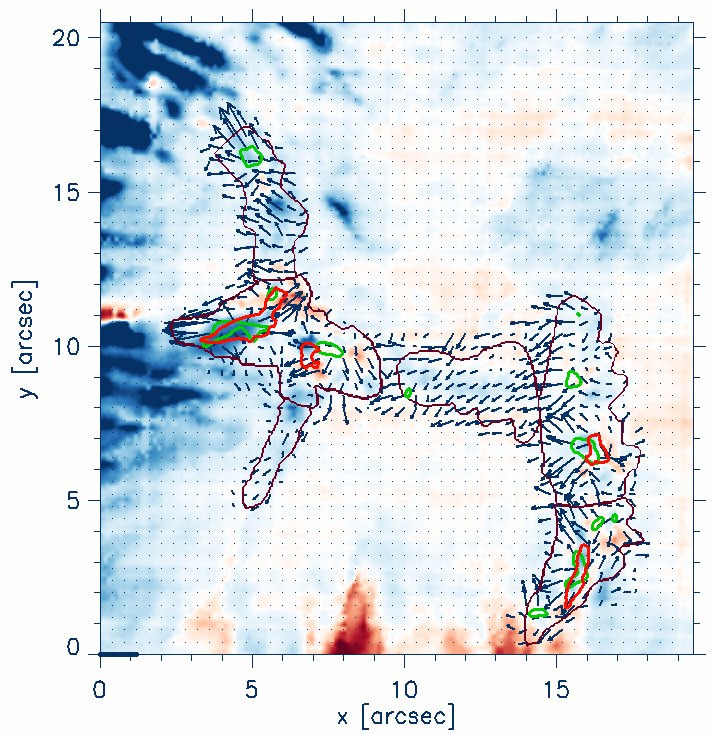}
\caption{Map of LOS velocities ($v_{\rm LOS}$) and horizontal LCT speeds ($\vec v_{\rm LCT}$). The colour coding of $v_{\rm LOS}$ is from $-2$~km\,s$^{-1}$ (blue) to 2~km\,s$^{-1}$ (red). Black arrows represent $\vec v_{\rm LCT}$; a black segment in the lower left corner corresponds to $v_{\rm LCT} = 1$~km\,s$^{-1}$. Green contours show centres of diverging motions and red contours delimit areas where an estimated plasma velocity is smaller than 1.5~km\,s$^{-1}$, which corresponds to the equipartition magnetic field of 340~G.
\label{fig:vlct}}
\end{figure}

A detailed map of LOS velocities at $\log \tau_{500\rm nm} = 0$ and apparent horizontal speeds of intensity structures (LCT speeds) is displayed in Fig.~\ref{fig:vlct}. The colour coding of $v_{\rm LOS}$ is the same as in Fig.~\ref{fig:maps}, that is, from $-2$~km\,s$^{-1}$ (blue) to 2~km\,s$^{-1}$ (red).
The horizontal LCT speeds ($\vec v_{\rm LCT}$) are represented by black arrows. Their magnitudes range from 0 to 1.2~km\,s$^{-1}$. They are probably larger, because the LCT method underestimates the magnitudes due to spatial and temporal averaging. Several centres of diverging motions are found: at the west and east of LB 1 with speeds of $\simeq 1$~km\,s$^{-1}$, at the south of LB~2 ($\geq 1$~km\,s$^{-1}$), two centres in LB~5a ($\simeq 0.7$~km\,s$^{-1}$), and one large in LB~5b ($\simeq 0.7$~km\,s$^{-1}$). A weak divergence is seen in LB~3 (speeds $< 0.3$~km\,s$^{-1}$) and no divergence but a slow unidirectional motion in LB~4. In Fig.~\ref{fig:vlct}, the divergence centres are marked by green contours, which represent the divergence $\nabla \cdot \vec v_{\rm LCT} = 0.08$~s$^{-1}$.

\subsection{Properties of LB granules}
\label{sec:gran}

\begin{figure}[]\centering
\includegraphics[width=0.38\textwidth]{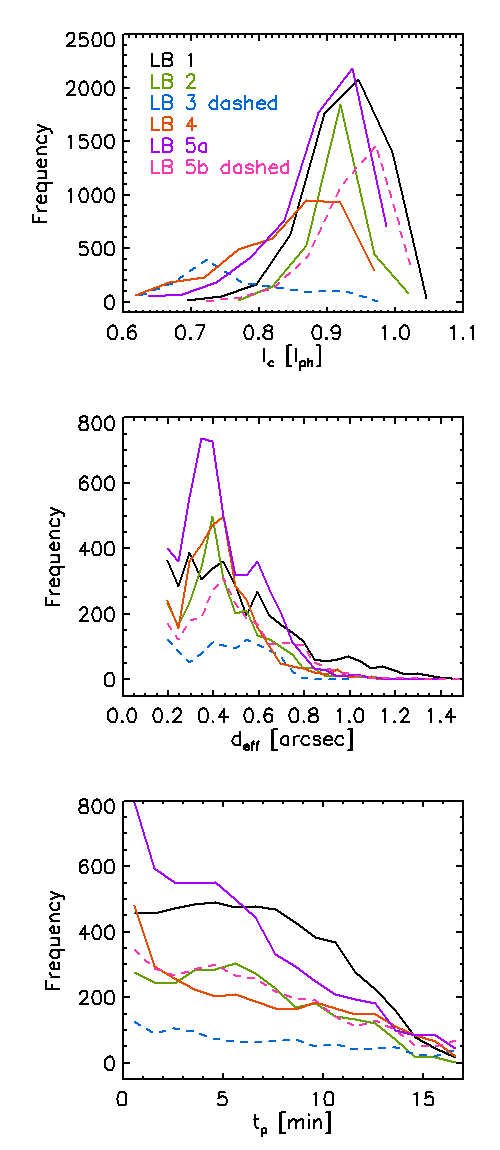}
\caption{Histograms of continuum intensities ($I_{\rm c}$) in units of the mean local quiet-Sun continuum $I_{\rm ph}$ at $\lambda = 1084$\,nm, effective diameters ($d_{\rm eff}$), and periods of presence ($t_{\rm p}$) of LB granules in individual LBs 1--5b.
\label{fig:hist2}}
\end{figure}

\begin{figure}[]\centering
\includegraphics[width=0.49\textwidth]{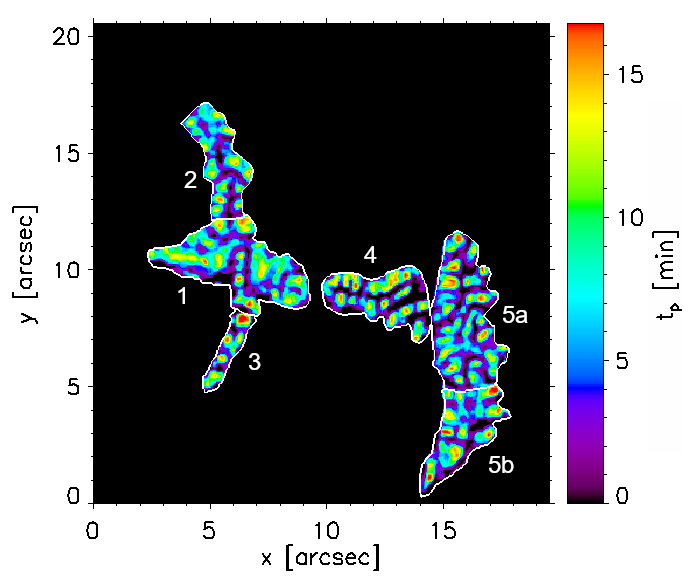}
\caption{Spatial distribution of presence periods (approximate lifetimes) of granules in LBs 1--5a.
\label{fig:lifmap}}
\end{figure}

All studied LBs consist of granules of different brightnesses, sizes, and lifetimes. Histograms of continuum intensities $I_{\rm c}$, effective diameters $d_{\rm eff}$, and periods of presence $t_{\rm p}$ that are a proxy of lifetimes (Sect.~\ref{sec:data}) are shown in Fig.~\ref{fig:hist2}. The intensity histograms are obviously similar to those of temperature in Fig.~\ref{fig:hist1}. The intensity range is from 0.6~$I_{\rm ph}$ (LBs 3 and 4) or 0.7~$I_{\rm ph}$ (other LBs) to 1.0--1.05~$I_{\rm ph}$ and the typical values are around 0.9--0.95~$I_{\rm ph}$ except for the weak LB~3. The effective diameters measured in the TiO image series are between 0\farcs 2 and 1\farcs 0 but large granules of $d_{\rm eff} \simeq$~1\farcs4 appear in LBs 1 and 5b. The most frequent $d_{\rm eff}$ is around 0\farcs 4. The histograms of presence periods show a trend of decreasing number of granules with an increasing period in all LBs. The mean presence period of LB granules is of about 6.2 minutes (Table~\ref{tab:ave}).
The spatial distribution of presence periods of LB granules in all LBs is shown in Fig.~\ref{fig:lifmap}. It is seen that the positions of small granules are quite stable, particularly in LBs~3 and 4 with minimum horizontal motions. Large continuous areas of $t_{\rm p} > 8$\,minutes in LB~1 correspond to large granules with $d_{\rm eff} > 1$\arcsec ~and $t_{\rm p}$ between 7 and 11 minutes. 

An inspection of the online TiO movie shows that two large granules appear at the beginning of the observation in the western part of LB~1, they grow in time, merge (frame 30), and then split into smaller granules (frame 60). They re-appear in frame 110 and are present till the end (frame 183). Their elongated shape is consistent with the inclined magnetic field \citep{Campos26}. The location of these granules is co-spatial with the centre of diverging motions at the position [5\arcsec, 10\arcsec] in Fig.~\ref{fig:vlct} indicating an expansion of the granules. An upflow of $-1.7$~km\,s$^{-1}$ is detected at this location. Another large expanding granule is formed in the eastern part of LB~1 (frames 37--140), also located near a divergence centre and an upflow area at the position [7\arcsec, 10\arcsec]. A similar situation is observed in the northern part of LB~5b -- two large granules appear in frames 10--40 and again in frames 120--183, in a region with a divergence centre and a weak upflow of $-0.6$~km\,s$^{-1}$ at the position [16\arcsec, 3\arcsec] in Fig.~\ref{fig:vlct}. The upflows and divergence centres in LB~2 at the position [5\arcsec, 16\arcsec] and LB~5a [16\arcsec, 7\arcsec] are not related to a formation of large granules.

The large granules are produced by an intense magnetoconvection \citep{Lagg14}. Comparing the densities of the magnetic and kinetic energy of the moving plasma,
\begin{equation}
    \frac{B^2}{8\pi} \approx \frac{1}{2}\,\rho v^2,
\end{equation}
we come to the concept of the equipartition magnetic field $B_{\rm eq} = \sqrt{4\pi \rho}\, v$, where $\rho$ denotes the mass density and $v$ is the characteristic plasma velocity. The balance of the magnetic and kinetic energy densities is important in layers deep below the visible surface but the observations and inversion method restrict us to make a rough estimate of the conditions at $\log \tau_{500\rm nm} = 0$, for which we have sufficient information. Near the large granules, the LOS velocities are in the range from $-1.7$ to 1.2~km\,s$^{-1}$ and the horizontal LCT speeds reach 1.0--1.2~km\,s$^{-1}$. These values may be underestimated due to the limited spatial resolution and LCT averaging. We also have to note that in the inversion model, the density is calculated assuming the hydrostatic equilibrium, which may be less accurate in the magnetised atmosphere as it is valid only if the magnetic field is aligned with the LOS. Adopting $v = 1.5$~km\,s$^{-1}$ and $\rho$ in the range of 3.1$\times 10^{-7}$--3.3$\times 10^{-7}$~g\,cm$^{-3}$, we obtain $B_{\rm eq} \simeq 340$~G. Because the magnetic field strength is weaker than this value in some areas in LBs~1 and 5b (cf. Figs.~\ref{fig:maps}, \ref{fig:hist1}, and \ref{fig:bzbh}), the convection in large granules is close to the equipartitional regime. In stronger magnetic fields, large expanding granules do not appear. To illustrate this, maps of $B$ and $\rho$ are used to calculate $v = B/\sqrt{4\pi \rho}$ and contours of $v = 1.5$~km\,s$^{-1}$ are plotted in red colour in Fig.~\ref{fig:vlct}. The delimited areas are co-spatial with the locations of large granules, centres of diverging horizontal motions, and upflows in LBs~1 and 5b.

\subsection{Statistical relations}
\label{sec:stat}

\begin{figure*}[]\centering
\includegraphics[width=\textwidth]{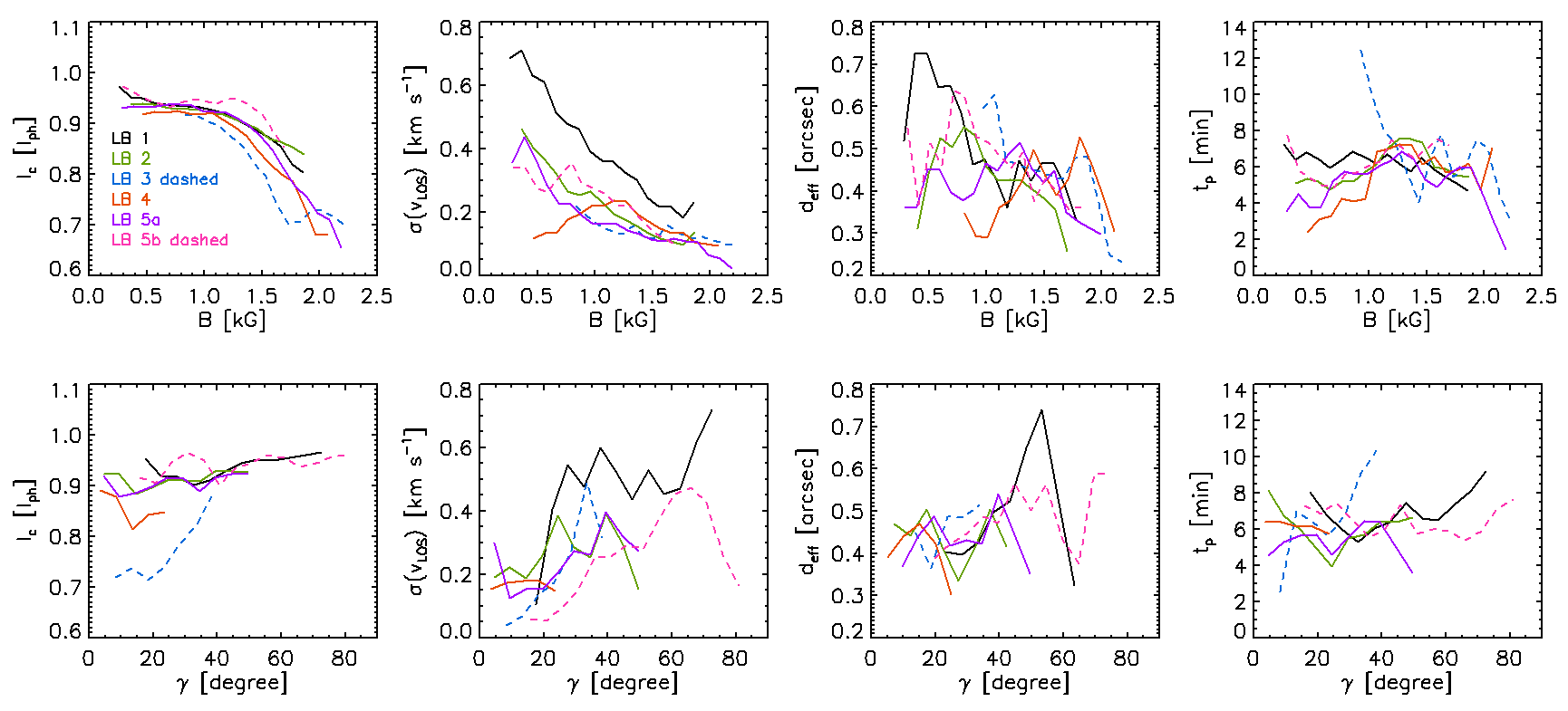}
\caption{Statistical relations of continuum intensity ($I_{\rm c}$) in units of the mean local quiet-Sun continuum $I_{\rm ph}$ at $\lambda = 1084$\,nm, scatter of the LOS velocity values ($\sigma (v_{\rm LOS})$), effective diameter ($d_{\rm eff}$), and period of presence ($t_{\rm p}$) of LB granules versus magnetic field strength $B$ (top row) and inclination $\gamma$ (bottom row) in individual LBs 1--5b. 
\label{fig:ivdl}}
\end{figure*}

The configuration of magnetic field is the main agent in the process of LBs formation and structure shaping. We study statistically the relations between the continuum intensity, standard deviation of the LOS velocity ($\sigma (v_{\rm LOS})$), effective diameter of LB granules, and their period of presence versus the magnetic field strength and inclination. For each LB separately, we calculate the mean $I_{\rm c}$, $\sigma (v_{\rm LOS})$, and $t_{\rm p} > 0$ at positions (pixels) where $B$ and $\gamma$ fall into 100~G and 5\degr ~bins of the $B$ and $\gamma$ histograms, respectively. Similarly, the mean values of $d_{\rm eff}$ are calculated in the same
bins using average $B$ and $\gamma$ in areas of individual segmented granules. This way we obtain mean values of the quantities under study corresponding to different values of magnetic field strength and inclination.
The results are shown in Fig.~\ref{fig:ivdl}. Statistical relations versus $B$ are plotted in the top row of the figure and those versus $\gamma$ in the bottom one. The standard deviations of the average values of $I_{\rm c}$, $\sigma (v_{\rm LOS})$, $d_{\rm eff}$, and $t_{\rm p}$ in the bins are typically 0.04~$I_{\rm ph}$, 100~m\,s$^{-1}$, 0\farcs 16, and 4~minutes, respectively.

The continuum intensity (Fig.~\ref{fig:ivdl}, first column) decreases with increasing $B$ in all LBs in a similar way. It is practically constant around 0.93~$I_{\rm ph}$ for $B < 1200$~G and then it decreases with an approximate rate of $-0.025~I_{\rm ph}$ per 100~G. The continuum intensity shows a weak ($0.01~I_{\rm ph}$ per 10\degr) and approximately linear increase with increasing (getting more horizontal) $\gamma$ in LBs~1, 2, 5a, and 5b. The narrow LB~3 is much darker than the other LBs and it reaches their brightness only in the neighbourhood with LB~1 (see Fig.~\ref{fig:bridges}), where $\gamma$ is large. This is the reason for the rapid increase of $I_{\rm c}$ with increasing $\gamma$. The small inclination range in LB~4 does not allow any trend to be determined. 

Because the most frequent values of LOS velocities are around zero (Fig.~\ref{fig:hist1}), we estimate the intensity of magnetoconvection using their scatter ($\sigma (v_{\rm LOS})$, Fig.~\ref{fig:ivdl}, second column). With the exception of LB~4, where only a weak upflow is present in the central channel with a weak magnetic field, all other LBs show a clear decrease of $\sigma (v_{\rm LOS})$ with increasing $B$. The approximate rates of decrease in the interval of 100~G are $-33$~m\,s$^{-1}$ for LB~1, $-19$~m\,s$^{-1}$ for LBs~2, 5a, 5b, and $-8$~m\,s$^{-1}$ for the narrow LB~3. The statistical relation between $\sigma (v_{\rm LOS})$ and magnetic inclination is quite noisy but a weak trend of increase with increasing $\gamma$ is seen for LBs~1, 3, 5a, 5b, and possibly~2. No trend is observed in LB~4 with an almost vertical field.

The relations of granular effective diameter (Fig.~\ref{fig:ivdl}, third column) to $B$ are different in individual LBs but there is a common trend of decreasing $d_{\rm eff}$ with increasing $B$ in LBs~1, 2, 3, and 5b. Low-inclination LBs~4 and 5a (mean $\gamma < 20$\degr), consisting of small granules with mean $d_{\rm eff} =$~0\farcs 43, do not show any clear relation between the size of granules and $B$. Concerning the relation with magnetic inclination, $d_{\rm eff}$ increases above the value of 0\farcs 5 only for $\gamma > 40$\degr ~in LBs~1 and 5b.

The periods of presence, which approximate the lifetimes of LB granules (Fig.~\ref{fig:ivdl}, last column), do not show any common trend with increasing $B$ and $\gamma$. The values of $t_{\rm p}$ are approximately constant with $B$ in LBs~1 and 5b, while in LBs~2, 4, and 5a they increase, reaching a peak of about 7~minutes at $B = 1300$~G and then decrease. In LB~3, the shape of the plot is influenced by a long-lived granule near a border with LB~1 (Fig.~\ref{fig:lifmap}), which is located in a weak and inclined magnetic field. We do not find any relation between $t_{\rm p}$ and $\gamma$ in the other LBs.

\section{Discussion and conclusions}
\label{sec:discuss}

We present an observation of five strong granular LBs located in a large complex sunspot. According to the HMI continuum images, most of the LBs were formed shortly (3--9 hours) before the observation. Because the sunspot was in the phase of growth, the LBs changed their characteristics quite fast. Their periods of life without substantial changes were in the range from 12 hours (LB~2) to 43 hours (LB~5). The values of temperature, magnetic field vector, and LOS velocity at the visible surface \mbox{($\log \tau_{500\rm nm} = 0$)} were obtained from SIR inversions of the infrared lines \ion{Si}{i} 1082.71\,nm and \ion{Ca}{i} 1083.90\,nm. The properties of LB granules, effective diameters, periods of presence, and apparent horizontal motions, were measured in a simultaneous series of images in the TiO band at $750.7 \pm 1.0$\,nm.

The temperatures, LOS velocities, magnetic field strengths, and inclinations reported in this work (Table~\ref{tab:ave} and Fig.~\ref{fig:hist1}) agree with the values stated previously in the literature -- we refer to the summary presented by \citet{Grinon21} in their tables~8 and 9. We confirm the well-known fact that the magnetic field in LBs is weaker than in surrounding umbrae. 
The magnetic inclination in LBs is influenced by an interaction of magnetic fields of neighbouring umbral cores, particularly of their horizontal components, which increase with radial distance from the core centres and meet in the areas of strong LBs (Fig.~\ref{fig:bzbh}). This results in a broad range of possible inclinations from zero to nearly horizontal and sometimes in the formation of magnetic canopies (LBs~1, 4, 5a).

The observed size of LB granules in the range of 0\farcs 2--1\farcs 4 with the most frequent value around 0\farcs 4 is in agreement with previously reported values \citep{SBV94, Hirzberger02, Falco16}. Their approximate lifetime (period of presence) of about 6 minutes might be underestimated due to the relatively short duration of the TiO series but it is similar to the result of \citet{Hirzberger02}. Apparent horizontal motions of the granules, detected by the LCT method, show slow unidirectional motions less than 0.5~km\,s$^{-1}$ or diverging motions of 0.7--1.2~km\,s$^{-1}$. 
A specific group are large granules with $d_{\rm eff} > $1\arcsec ~in LBs~1 and 5b. These granules expand, which is confirmed by the centres of diverging motions at their location and are co-spatial with upflows from $-0.6$ to $-1.7$~km\,s$^{-1}$. Their local magnetic field strength, approximately 300~G, is comparable with the equipartition field (340~G, Sect.~\ref{sec:gran}) estimated from the local model atmosphere.

Although each of the observed LBs has its own individual characteristics, we found several relatively common rules regarding the influence of the magnetic field strength on the continuum brightness, LOS velocities, and sizes of LB granules (Fig.~\ref{fig:ivdl}). The brightness, which is comparable to the brightness of the penumbra at a weaker magnetic field, begins to decrease significantly only at $B >$ 1200~G. The dispersion of LOS velocities clearly decreases with increasing $B$. The effective diameter of LB granules decreases with increasing $B$ but LBs with a strong and nearly vertical magnetic field have only small granules ($d_{\rm eff} \simeq $ 0\farcs 4) and do not show this trend.

Our observations confirm the magnetoconvective origin of strong granular LBs. These structures represent a transition between the quiet Sun non-magnetic overturning convection and an oscillatory convection \citep{Weiss90, Schussler06} in a strong ($> 2$~kG) vertical magnetic field of the umbra. The latter type of magnetoconvection takes place in narrow field-aligned columns manifesting as umbral dots. A local weakening of magnetic field to typically 1~kG and less, caused either by a penetration of hot plasma from deep layers \citep{Rempel11} or by a compression of photospheric granulation and parts of penumbra during a sunspot evolution (which was probably our case) results in the expansion of convective morphology from restricted umbral-dot scales through small static LB granules to growing cells similar to quiet-Sun granules. The observation that the largest granules ($ >1$\arcsec) emerge in areas with $B$ near the equipartition threshold indicates a localized regime where the kinetic energy density ($1/2\,\rho v^2$) balances or exceeds the magnetic energy density ($B^2 / 8\pi$). 

The statistical inverse relationship between the field strength and continuum intensity, LOS velocity dispersion, and granular diameter is a direct manifestation of that process. Where $B$ is higher than the equipartition threshold ($ B \simeq 340$~G), the Lorentz force restricts horizontal gas expansion and suppresses the overturning convective motion. This also applies to whole active regions and surrounding quiet photosphere, where the brightness, size, shape, and LOS velocity of convective cells are limited by the magnetic field vector, as demonstrated, for example, by \citet{Campos26}. 
We have to note that the magnetic field inclination in the granular LBs under study is not extremely large, typically 15\degr --40\degr ~and smaller than 60\degr ~in most cases. This causes an elongated shape of some large granules as a maximum effect (Sect.~\ref{sec:gran}). A highly inclined magnetic field with a strong horizontal component would lead to a distinct, horizontal mode of magnetoconvection, present, for example, in penumbral filaments.

\begin{acknowledgements}

We thank the anonymous referee for valuable comments, which helped us to improve the paper. This work was supported by the Czech-German common grant, funded by the Czech Science Foundation under the project 23-07633K and by the Deutsche Forschungsgemeinschaft under the project BE 5771/3-1 (eBer-23-13412), and the institutional support ASU:67985815 of the Czech Academy of Sciences. GREGOR observations were supported by SOLARNET project that has received funding from the European Union Horizon 2020 research and innovation programme under grant agreement no. 824135.
The \mbox{1.5-metre} GREGOR solar telescope was built by a German consortium under the leadership of the Institut f\"ur Sonnenphysik (KIS) in Freiburg with the Leibniz-Institut f\"ur Astrophysik Potsdam, the Institut f\"ur Astrophysik Göttingen, and the Max-Planck Institut f\"ur Sonnensystemforschung in G\"ottingen as partners, and with contributions by the Instituto de Astrof\'\i sica de Canarias and the Astronomical Institute of the  Czech Academy of Sciences.
HMI data are courtesy of NASA/SDO and the HMI science team.

\end{acknowledgements}


\bibliographystyle{aa}
\bibliography{bibliography3}


\end{document}